\documentclass[aps,floatfix,amsmath,amssymb,reprint,superscriptaddress]{revtex4-2}

\usepackage{graphicx}
\usepackage{dcolumn}
\usepackage{bm}
\usepackage[colorlinks=True,allcolors=blue]{hyperref}
\usepackage{newtxtext,newtxmath}
\usepackage{microtype}
\usepackage{orcidlink}
\usepackage{lipsum}
\usepackage{amsmath,amssymb}
\usepackage{ulem} 
\usepackage{pdfpages}
\usepackage{siunitx}
\usepackage{sidecap}
\usepackage{mhchem} 
\usepackage[detect-all]{siunitx}

\DeclareSIUnit{\molar}{M}

\makeatletter 

\AtBeginDocument{\let\LS@rot\@undefined}
\makeatother
\makeatletter
\renewcommand\@make@capt@title[2]{%
 \@ifx@empty\float@link{\@firstofone}{\expandafter\href\expandafter{\float@link}}%
  {\textbf{#1}}\@caption@fignum@sep#2\quad
}%
\makeatother

\newcommand{\kB}{k_{\text{B}}}
\newcommand{\kT}{\kB T}

\newcommand{\leibnizd}[1]{\mathrm{d}{#1}}

\newcommand{\UCP}{U_{\mathrm{CP}}}
\newcommand{\UEP}{U_{\mathrm{EP}}}
\newcommand{\UDP}{U_{\text{DP}}}
\newcommand{\UDO}{U_{\text{DO}}}

\newcommand{\Upol}{U_{\text{pol}}}

\newcommand{\Eq}[1]{Eq.~\ref{#1}}
\newcommand{\Eqs}[1]{Eqs.~\ref{#1}}
\newcommand{\Fig}[1]{Fig.~\ref{#1}}
\newcommand{\partFig}[2]{Fig.~\hyperref[#1]{\ref*{#1}#2}}
\newcommand{\partFigure}[2]{Figure~\hyperref[#1]{\ref*{#1}#2}}

\newcommand{\lu}{Leiden Institute of Physics (LION), Leiden University, Niels Bohrweg 2, 2333 CA Leiden, the Netherlands}
\newcommand{\ue}{SUPA, School of Physics and Astronomy, University of Edinburgh, Peter Guthrie Tait Road, Edinburgh EH9 3FD, United Kingdom}
\newcommand{\hc}{The Hartree Centre, STFC Daresbury Laboratory, Warrington WA4 4AD, United Kingdom}

\begin{document}

\title{Self-Generated Electric Fields in Polyelectrolyte Gradients Increase Microparticle Transport}

\author{Max Huisman}
    \thanks{These authors contributed equally to this work.}
    \affiliation{\lu}
\author{Ali Azadbakht}
    \thanks{These authors contributed equally to this work.}
    \affiliation{\lu}
\author{Patrick B. Warren}
    \affiliation{\ue}
    \affiliation{\hc}
\author{Daniela J. Kraft}
    \affiliation{\lu}

\begin{abstract}
    There are many situations in nature and industry where small particles are exposed to gradients of charged polymers, such as enzymes in biological gradients of DNA or RNA, virus particles in respiratory droplets, and colloidal particles in stratifying paint layers. Here, we study the phoretic propulsion of charged microparticles in a polyelectrolyte gradient. We theoretically predict the emergence of a macroscopic electric field from charge-separation dynamics in a polyelectrolyte gradient under a continuous diffusive driving force. We confirm the presence of this self-generated electric field experimentally and show that it significantly increases the phoretic velocity of the microparticles. Finally, for high molecular weight polyelectrolytes we observe that propulsion becomes gradient-independent, consistent with diffusiophoretic predictions for asymmetric electrolytes. Our results show that self-generated electric fields in polyelectrolyte gradients can enhance microparticle transport, with potential applicability wherever charged species of different mobility are continuously driven out of equilibrium.
\end{abstract}

\keywords{electrophoresis, charge-separation dynamics, macromolecular gradient, diffusiophoresis, particle transport}

\maketitle



\section{Introduction}


Recent respiratory virus pandemics, including SARS-CoV-2 and influenza, have highlighted the importance of understanding dynamics within virus-containing respiratory droplets. Studies \cite{Huynh2022,Oswin2022,Iida2026} show that during evaporation of such droplets, the advective push of the receding interface can result in steep concentration gradients of mucin, a complex group of charged macromolecules and the droplet's main polymeric compound. Similar gradient formation occurs in coatings, where charged macromolecules are often used for added stability through charge complexation \cite{Argaiz2023}, yet during the drying process steep concentration gradients inevitably form \cite{Okuzono2006,Huisman2025}.

The uneven distribution of charged macromolecules and counterions, combined with the evaporative driving force, may lead to the formation of a macroscopic electric field $E$. The mechanism is intuitive: when macromolecules and counterions 
have different mobilities, the two species separate, and an electric field forms to oppose this separation and preserve charge neutrality. At thermodynamic equilibrium this effect is well established, for instance in the inflated sedimentation height of charged colloids under gravity \cite{Racsa2004,Philipse2004}. Whether the same holds in non-equilibrium processes like the evaporation of respiratory droplets or the drying of paint, where the advective driving force continuously transports macromolecules and counterions of differing mobility, is far less obvious.

An electric field can directly affect the dynamics of dispersed particles with a surface charge. Small particles in solute gradients are expected to move through chemiphoresis \cite{Anderson1989, Ault2025, Shi2025}, but at the surface of a charged particle an additional slip velocity is generated if an electric field is present, leading also to electrophoretic motion. Electrophoretic motion of negatively charged RNA viruses inside respiratory droplets could modulate viral viability by moving them towards the interior of the droplet, where they are shielded from destabilization by a dense polymer layer at the solution-air interface \cite{Brackley2021,Pease2022,Huisman2023}. For industrial applications, electrophoresis in polyelectrolyte gradients offers an additional handle for precise control on stratification morphology in colloid-polymer mixtures \cite{Fortini2016,Sear2017,Sear2018} and colloidal focusing in microfluidics applications \cite{Shin2016,Ault2017}.

\begin{figure*}[t]
\includegraphics[width=0.95\textwidth]{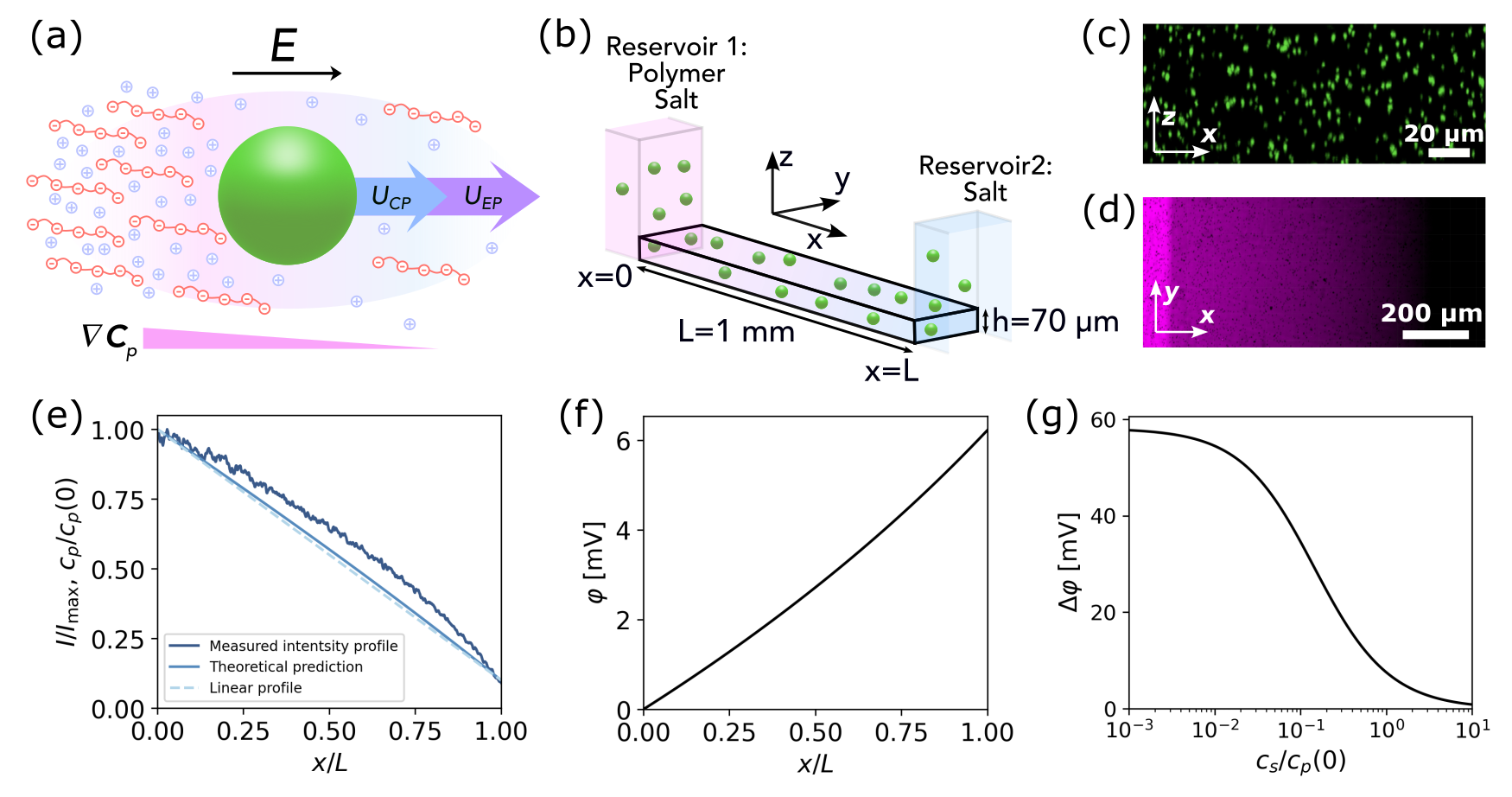}
\caption{\label{fig:1} A gradient of diffusing polyelectrolyte leads to the formation of an electric field. (a) Schematic of a negatively charged colloid in a gradient of polyelectrolyte. Additive contributions of electrophoresis ($\UEP$) and chemiphoresis ($\UCP$) propel a particle down the polyelectrolyte gradient. (b) Schematic of the experiment, which is a limited-diffusion liquid junction where two reservoirs are connected by a narrow channel. All reservoirs contain the same concentration of colloids and salt, while polymer is added to only one of the two reservoirs. (c) $y$-axis maximum intensity projection, showing a snapshot of fluorescent polystyrene particles in the microchannel. (d) Confocal microscopy image over the entire width of the channel, showing an intensity gradient of the rhodamine B signal to the PSS polymers. Image was taken 2 hours after the start of the experiment. (e) Comparison between the averaged intensity profiles calculated from (d) and the theoretically predicted polymer concentration profiles from \Eqs{eq:BVP_our_system}. (f) Calculated electrostatic potential $\varphi(x)$ over the channel, in a system with $c_s=\SI{150}{\milli\molar}$. (g) Calculated liquid junction potential across the channel $\Delta \varphi$ with $c_s/c_p(0)$. For this plot we used the ratio $c_p(0) : c_p(L) = 10 : 1$. We observe that the junction potential $\Delta \varphi \simeq 2.2\,\kT/e\simeq\SI{56}{\milli\volt}$ in the unscreened limit where $c_s/c_p(0) \rightarrow 0$ while $\Delta \varphi \rightarrow 0$ in the charge-screened limit where $c_s/c_p(0) \rightarrow 10$. The units for $\varphi$ have been converted from $\kT/e$ to \unit{\milli\volt} to facilitate later analyses.} 
\end{figure*}

In this work, we show that a polyelectrolyte gradient, sustained by a continuous diffusive flux, leads to a macroscopic electric field $E$ that increases microparticle transport (\partFig{fig:1}{a}). We mimic non-equilibrium conditions through a quasi-steady state concentration gradient using a simple restricted diffusion liquid junction. In this setup two large reservoirs are connected by a narrow channel, and one of the reservoirs contains polyelectrolyte while the other one does not. We first present a theoretical analysis for the steady state solutions of the concentration profiles of ionic species in the channel. We find that the non-zero ion flux through the channel is compensated by an electrostatic potential difference $\Delta \varphi$ over the channel which establishes an electric field. We then show experimentally that this electric field enhances the phoretic propulsion speed of negatively charged colloids in a polyelectrolyte gradient compared to the neutral polymer. Finally, we observe different dynamics for very high molecular weight polyelectrolytes, where the propulsion rate does not depend on the gradient size.

\section{Concentration Profiles and Electrostatic Potential in the Channel}
We experimentally study the motion of negatively charged colloids with diameter \SI{1.25}{\micro\metre} in a polymer gradient by employing a setup consisting of a narrow channel connecting two large reservoirs where only one of the reservoirs contains polymer or polyelectrolyte, see Methods for a detailed description. The reservoirs are large compared to the channel size, resulting in a quasi-steady concentration gradient in the channel that mimics the non-equilibrium conditions inside evaporating respiratory droplets. At the same time, the dynamics in the channel are sufficiently slow that we can treat the system as approximately steady state for theoretical comparison. 

To test for the formation of a gradient across the channel we experimentally measured the concentration profile of poly(sodium 4-styrenesulfonate) (NaPSS) in the channel, using NaPSS $M_W=\SI{70}{\kilo\dalton}$ stained with rhodamine B; see Methods. We take measurements after an initialization period of 60 minutes, after which time the manufacturer suggests that a linear solute gradient can be observed, that persists for up to a few days. As such, we can treat the system as a steady state. A confocal microscopy image of the entire channel in \partFig{fig:1}{d} clearly shows the gradient in the signal of rhodamine B. If we calculate the average intensities we observe that the profile does not deviate far from the linear profile that is expected in the channel, \partFig{fig:1}{e}. 

Assuming steady state conditions, we proceed to calculate the distribution of the charged solutes in the channel and, importantly, show how this affects the electrostatic potential $\varphi$ across the channel. At early times after the experiment is set up, there is a concentration gradient for the polymer and, as a consequence, counterions, and a flat concentration profile for the co-ions. In the experiments that we will describe below, we consider the distribution of the polyelectrolyte NaPSS in the presence of the salt NaCl, corresponding to poly(styrenesulfonate) (PSS-) polymers, sodium (Na+) counterions and chloride (Cl-) co-ions. Given that the polymer diffusivity is essentially negligible compared to the ion diffusivities, this would lead to a net flux of positive counterions diffusing from the polymer-rich compartment into the polymer-poor compartment. Such a flux would lead to a charge build up in the two compartments: negative in the polymer-rich compartment and positive in the polymer-poor compartment. This charge imbalance would in turn generate an electric field along the capillary, retarding the counterion flux and driving a flux of co-ions. In the final steady state, the electrostatic potential gradient along the capillary is such that the net ion current vanishes. 

To analyse this problem, we use the Nernst-Planck equations that describe the ion fluxes in the presence of concentration gradients and an electric field, seeking a solution which satisfies the boundary conditions for the polymer and ion concentrations, and has a vanishing net ion current. In this case a closed-form solution can be obtained, as described in the Supporting Information (SI), with coefficients determined by solving a non-linear equation set derived from the boundary conditions. The system of ODEs that underpins this reads
\begin{subequations}\label{eq:BVP_our_system}
\begin{align}
    \frac{\partial \varphi}{\partial x} & = \frac{1}{2(c_p + c_s)} \left[\left(1-\frac{D_p}{D_c} \right)J'_p +\left(1-\frac{D_s}{D_c} \right)J'_s \right],  \\
    \frac{\partial c_p}{\partial x} &= -J_p' + c_p \frac{\partial \varphi}{\partial x}\,,~~~ 
    \frac{\partial c_s}{\partial x} = -J_s' + c_s \frac{\partial \varphi}{\partial x}\,, \tag{\ref*{eq:BVP_our_system}b, \ref*{eq:BVP_our_system}c}
\end{align}
\end{subequations}
where $J_i'$ is a reduced molar flux of species $i$, $D_i$ the diffusion coefficient [\unit{\metre\squared\per\second}], $c_i$ the charge concentration in [\unit{\mole\per\meter\cubed}], $\varphi$ the electrostatic potential in [$\kT/e$] and subscripts are used for the polymer ($p$), counterion ($s$) and co-ion ($c$). 

\begin{figure*}[t]
\includegraphics[width=0.9\textwidth]{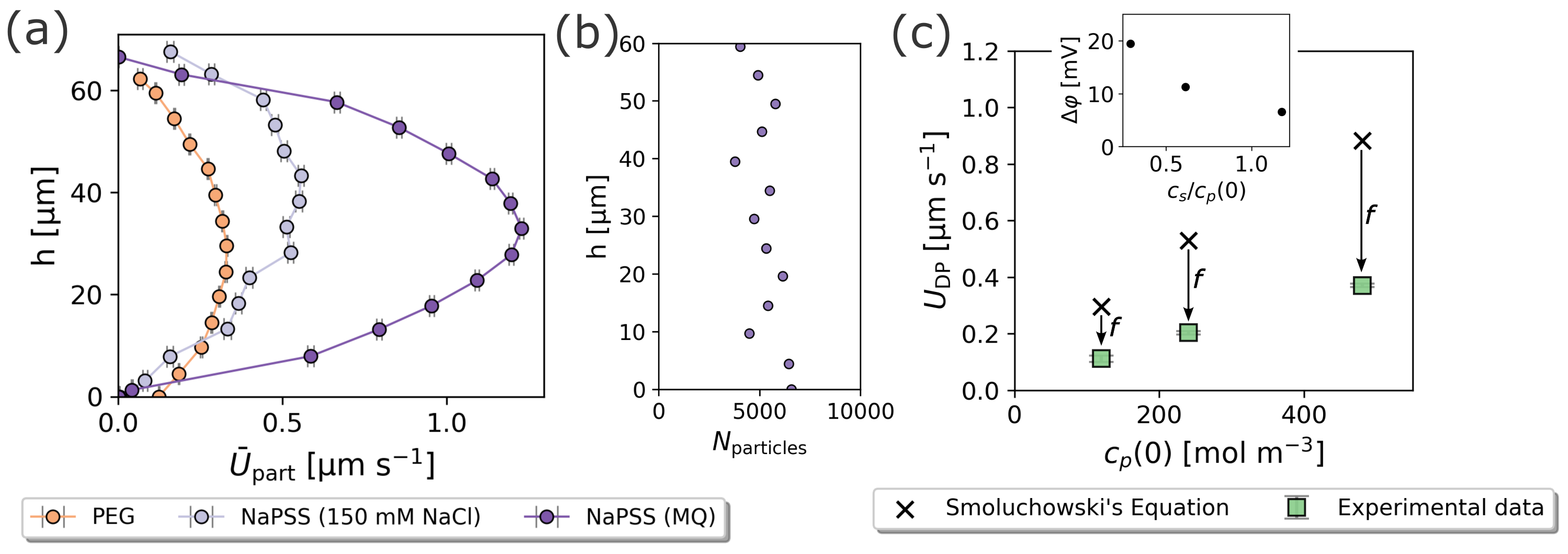}
\caption{\label{fig:2} Electric field in polyelectrolyte gradients enhances microparticle transport. (a) Experimentally measured velocities of negatively charged polystyrene colloids (\ce{-COOH} endgroups, $R = \SI{1.25}{\micro\metre}$) in gradients of neutral polymer (PEG) and polyelectrolyte (NaPSS, with and without added salt). The polymer concentrations in the left reservoir is 10 wt\% for all systems. (b) Number of particles for which velocities were measured over the duration of the experiment per height bin for PEG in panel (a). (c) Diffusiophoretic drift velocities $\UDP$ as a function of polymer concentration in the left reservoir $c_p(0)$, for NaPSS 70k systems with \SI{150}{\milli\molar} background salt. We calculated $\UDP = \frac{2}{3} U_x(z=\frac{h}{2})$ from parabolic velocity profiles such as those presented in panel (a). Plotted alongside, black crosses, are predictions from \Eq{eq:Smoluchowski} for electrophoresis due to the electrostatic potential across the channel as presented in \partFig{fig:1}{g}. Black arrows are used to indicate how the order of magnitude of theoretical predictions can be reduced by including the effective charge per monomer $f$ into the expression of \Eq{eq:Smoluchowski}. Inset shows the variation of the electrostatic potential across the channel $\Delta \varphi$ for the ratios of background salt and left reservoir concentration that we used in experiments: $c_s/c_{p}(0) \simeq~$1.2, 0.6 and 0.3, respectively.} 
\end{figure*}

Since we are only concerned with gradients in the electrostatic potential in \Eqs{eq:BVP_our_system}, the system is unaffected by a constant shift in $\varphi$. Therefore, the liquid junction potential can be expressed as $\Delta \varphi = \varphi(L)$, where $\varphi(0) = 0$ and we initially set $\varphi=0$ throughout the channel. Furthermore, we initialize the system for a typical experimental situation with boundary conditions $c_p(0)=\SI{120}{\mole\per\meter\cubed}$ and $c_p(L)=0.1 \times c_p(0) = \SI{12}{\mole\per\meter\cubed}$, as we observed that the experimentally measured fluorescence intensity at $x=L$ reduces to about 10\% of its maximum value in \partFig{fig:1}{e}. The salt is evenly distributed such that $c_s(0) = c_s(L)=\SI{150}{\milli\molar}$. For the diffusion coefficients we use $D_s = \SI{1.33e-9}{\metre\squared\per\second}$ (Na$^+$), $D_c = \SI{2.03e-9}{\metre\squared\per\second}$ (Cl$^-$) and for the polyelectrolyte we use that ${D_p}/{D_c} \ll 1 \simeq 0$, which still gives a converging solution to the boundary value problem (BVP). 

\partFig{fig:1}{e} shows the calculated theoretical concentration profile of the polymer, which is near-linear (striped lines), as advertised. We note that is also possible to obtain an exact solution to the system of differential \Eqs{eq:BVP_our_system}, as shown in the SI, with identical results. We observe a slight positive offset in the concentration profiles of the polymer at lower salt (see SI) that is compensated by a decreased co-ion concentration.
These theoretical predictions agree very well with our experimental measurements using fluorescently labelled NaPSS as shown in \partFig{fig:1}{e}. The measured concentration profile also displays a positive deviation from linearity, which is slightly more strongly expressed than in the theoretically predicted profile.

Although $\partial\varphi/\partial x$ could be eliminated in the system of \Eqs{eq:BVP_our_system}, by retaining it we are able to calculate the electrostatic potential over the channel, see \partFig{fig:1}{f}. We observe that $\varphi(x)$ is an increasing function of $x$, meaning that the electric field $E = -{\partial \varphi}/{\partial x}$ is directed down the polymer concentration gradient. For our system initialized at $\varphi(x) = 0$, the potential difference over the channel $\Delta \varphi =  \varphi(L)$. $\Delta \varphi$ arises from the ion current $I = \Sigma_i z_i J_i$, which is in general non-zero for some $\Delta \varphi$ as there are constant fluxes $J_i$ of charged species through the channel (see above). There is no return path for this current, such as another channel or electrochemical reactions. Therefore, charge builds up in the reservoirs, until it balances the ion current and $I \rightarrow 0$ for a specific value $\Delta \varphi$ in the steady state.

This electrostatic potential difference $\Delta \varphi$ is furthermore sensitive to the ratio of added salt compared to the polymer concentration in the left reservoir, $c_s/c_p(0)$. We show this dependence in \partFig{fig:1}{g} for a polymer concentration ratio $c_p(0)$\,:\,$c_p(L) = 10$\,:\,1, which was chosen to match the experimentally observed intensity difference from \partFig{fig:1}{e}. In the low-salt limit $\Delta \varphi \simeq \SI{56}{\milli\volt}$ while in the high-salt limit $\Delta \varphi \rightarrow 0$, as we expect for a charge-screened system.

\section{Diffusiophoretic propulsion and velocity profiles}
\label{sec:DP_profiles}

To elucidate particle transport in gradients of charged polyelectrolyte, we start by testing microparticle motion in a gradient of neutral polymer (polyethylene glycol, PEG) as this allows us to characterize diffusiophoretic transport without electrostatic effects. We experimentally measured the velocities of negatively charged polystyrene colloids at $x\simeq L/2$ (\partFig{fig:2}{a}); see Methods. The particles were homogeneously distributed in the starting system and all solutions were density matched, such that buoyancy flows were negligible (see SI). After leaving the system to settle for about 60 minutes, we observe that in a PEG gradient all colloids move unidirectionally from high to low polymer concentration (\partFig{fig:2}{a}), with similar amounts of particles in every height bin (\partFig{fig:2}{b}). 

In the case of the neutral polymer (PEG), propulsion results from the size exclusion interaction between the polymer solution and the surface of the colloids as suggested in \cite{Sear2017}. The particles are also not driven by the polymer flux through the channel since $\Upol\simeq\SI{1}{\nano\metre\per\second}$ (see SI), which is about 1000\,$\times$ slower compared to the movement of the particles at $\sim \SI{1}{\micro\metre\per\second}$. Instead, the parabolic velocity profile in \partFig{fig:2}{a} can be explained by considering the flow field in the channel, that results from the balance between diffusioosmosis and planar Pousseuille flow, as typically found in closed-channel geometries \cite{Shin2016}. The flow profile is found by solving the Stokes equation for $U_x(z)$
with boundary conditions $U_x(z=0) = U_x(z=h) = \UDO$ (diffusioosmotic slip velocity) and satisfying the condition of no net flow, $\int^h_0\leibnizd{z}\, U_x(z) = 0$, which should hold for the closed system. One finds that the \emph{net} drift speed of the particles can be described by
\begin{equation}
    \label{eq:parabolic_profile}
    U_{\text{part}}(z) = \UDP + \UDO \left[1-6 z \frac{(h-z)}{h^2}\right],
\end{equation}
where $\UDP$ is the diffusiophoretic drift velocity of the particles relative to the fluid. Since both $\UDP$ and $\UDO$ are independent of height, \Eq{eq:parabolic_profile} captures the observed parabolic profile. 

Investigating more closely the velocity profile in a PEG gradient (\partFig{fig:2}{a}), the particle velocities at the top and bottom of the channel do not completely vanish. According to \Eq{eq:parabolic_profile}, this suggests a mismatch between diffusiophoresis and diffusioosmosis. Alternatively these dynamics may result from finite size effects, if the size of the PEG molecules becomes comparable to the thickness of the particles' boundary layer \cite{Shin2016}. 

\section{Electrophoresis increases particle velocities}

Next, we investigate whether the emergence of an electrostatic field across the channel, predicted to arise in a polyelectrolyte gradient (\partFig{fig:1}{g}), enhances microparticle phoretic speeds. To that end, we replace the neutral polymer (PEG) by a polyelectrolyte (NaPSS). As shown in \partFig{fig:2}{a}, when using the same mass concentration gradients we observe the same directionality of movement, with particles moving down the gradient. This is in agreement with recent work using the same polymers to investigate the exclusion distance of particles in a dead-end pore geometry \cite{Akdeniz2024}.

For polyelectrolyte gradients in \partFig{fig:2}{a} we consistently observe that $U_{\text{part}}(z=0) =U_{\text{part}}(z=h) \simeq 0$. Considering the boundary conditions for the fluid flow profile, $U(z=0) = U(z=h) = \UDO$, we infer that the particle's diffusiophoretic drift velocity $\UDP \simeq - \UDO$ throughout the system. For this case, we obtain from \Eq{eq:parabolic_profile} that the maximum velocity in the middle of the channel $U_{\rm{part}}(z=\frac{h}{2}) = \frac{3}{2} \UDP$, allowing us to directly extract the phoretic velocity from the experimentally measured velocity of the tracer particles in the channel.

Moreover, in the NaPSS gradient at 150 mM background salt (NaCl) the particle velocity increases by about a factor 2 compared to the PEG gradient (\partFig{fig:2}{a}). This observation suggests that the contributions of chemiphoresis due to the chemical gradient and electrophoresis due to the electric field are additive, and we define the overall diffusiophoretic velocity as $\UDP = \UCP + \UEP$.

\begin{figure*}[t]
\includegraphics[width=0.9\textwidth]{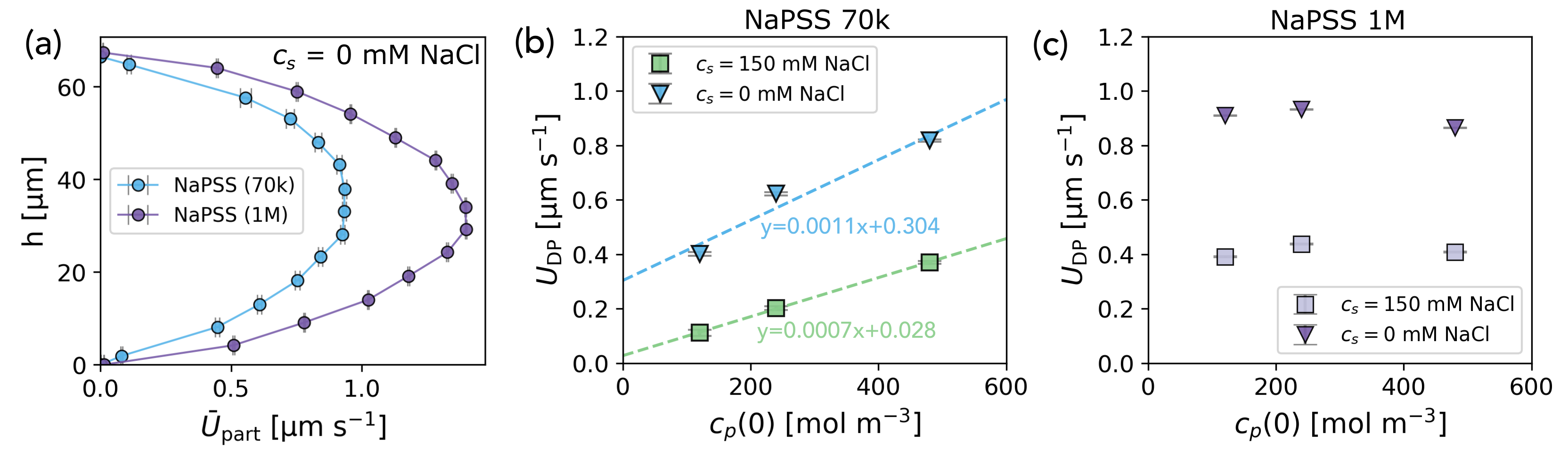}
\caption{\label{fig:3} Gradients under low salt conditions increase phoretic propulsion speed, and high molecular weight NaPSS changes the dominant diffusiophoretic mechanism. (a) Comparison between the velocity profiles in a system with NaPSS 70k and NaPSS 1M. In both cases, the reservoir concentration is 5\,wt\% and there is no salt added to the system. (b) For NaPSS 70k, comparison between the diffusiophoretic drift velocity $\UDP$ in a system with $c_s=\SI{150}{\milli\molar}$ and without added salt and $c_s/c_p \ll 1$ as a function of the polymer concentration in the left reservoir $c_p(0)$
$\UDP$ is increased for the low salt system; dashed lines are a linear fit to the data (the fit parameters are annotated) and remain similar for both cases. 
(c) In contrast, for high concentrations of polyelectrolyte (NaPSS 1M), measured values of $\UDP$ for various polymer concentrations in the left reservoir $c_p(0)$ remain approximately constant in systems with and without added salt, while $\UDP$ is again increased for the low-salt system.} 
\end{figure*}

For the chemiphoretic contribution, the propulsion mechanism from size exclusion (depletion) results in $\UCP \sim R^2\, {\partial c_p}/{\partial x}$ \cite{Anderson1989, Sear2017}. For NaPSS 70k at reservoir concentration 10\,wt\%, the gradient ${\partial c_p}/{\partial x}$ reduces by a factor 2 compared to PEG 35k at reservoir concentration 10\,wt\%. Using the Flory scaling $R_g \sim b N^{0.588}$ for a good solvent, which we assume is approximately valid for NaPSS at our experimental conditions with added salt, and where $b$ is the monomer size and $N$ is the degree of polymerization, we find that $R_{g, \rm{PEG}}^2 \gtrsim 2 R_{g, \rm{NaPSS}}^2$. This result suggests that $\UCP$ in a PEG gradient is similar or faster compared to in a NaPSS gradient, and therefore that the electric field induced electrophoresis is responsible for the increased value of $\UDP$.

To estimate whether it is realistic that the electrophoretic contribution $\UEP$ significantly increases the particle velocities, we consider the steady state electrostatic potential difference across the channel $\Delta \varphi$. For 10\,wt\% NaPSS with \SI{150}{\milli\molar} background salt $c_s/c_{p}(0) \simeq 0.3$ and we find from \partFig{fig:1}{g} that $\Delta \varphi \simeq \SI{20}{\milli\volt}$. This potential difference results in an electric field of strength $E = -{\partial \varphi}/{\partial x} \simeq -{\Delta \varphi}/{L} \simeq -\SI{20}{\volt\per\metre}$, where we used the channel length $L=\SI{1}{\milli\metre}$. During electrophoresis of a solid charged particle in an electric field, the relation between particle velocity and the electric field strength is given by Smoluchowski's equation
\begin{equation}
    \label{eq:Smoluchowski}
    \UEP = \frac{\varepsilon_r \varepsilon_0 \zeta}{\eta}\times E,
\end{equation}
where $\varepsilon_r$ is the dielectric constant of water, $\varepsilon_0$ the vacuum permittivity, $\zeta$ the surface potential of the particle, $\eta$ the viscosity of the surrounding fluid and $E = -{
\partial\varphi}/{\partial x}$ the electric field in the $x$-direction. An electric field of strength $E \simeq- \SI{20}{\volt\per\metre}$ can thus propel a particle with a surface charge $\zeta = -\SI{100}{\milli\volt}$ through water at a velocity $\UEP \simeq\SI{1}{\micro\metre\per\second}$, similar to the velocities observed in \Fig{fig:2}, and down the polyelectrolyte gradient. Therefore, it appears reasonable that electrophoresis explains the increased particle velocities in the NaPSS gradient.

To assess how the size of the electric field affects the total diffusiophoretic drift velocity $\UDP$ in our experiments, one approach is to vary the polymer reservoir concentrations $c_{p}(0)$ at constant salt $c_s$, which according to \partFig{fig:1}{g} changes the steady state potential difference across the channel $\Delta \varphi$. We calculate $\UDP = \frac{2}{3} U_x(z=\frac{h}{2})$ from our experimental data at various $c_{p}(0)$. For $c_s = \SI{150}{\milli\molar}$ we observe that $\UDP$ increases with $c_p(0)$, and the trend appears linear as highlighted using a linear fit (dashed green line), see \partFig{fig:2}{c}. A linear variation of $\UDP$ with $c_p(0)$ is consistent with the hypothesis that $\UCP$ and $\UEP$ are additive. 

For a non-adsorbing polymer, diffusiophoretic velocity scales linearly with the polymer concentration gradient, as shown in \cite{Sear2017}. We suggest that it is correct to employ this model, since the negatively charged colloids and polyelectrolyte repel each other, as was also assumed in \cite{Akdeniz2024}. Furthermore, according to \Eq{eq:Smoluchowski}, $\UEP \simeq\Delta \varphi$, while $\Delta \varphi$ changes with $c_s/c_p(0)$ in the reservoir, as shown in \partFig{fig:1}{g}. The reservoir concentrations in \partFig{fig:2}{c} corresponds to $c_s/c_{p}(0) \simeq~$1.2, 0.6, and 0.3, respectively. Under these conditions $\Delta \varphi$ decreases with $c_s/c_p(0)$, inset of \partFig{fig:2}{c}, and the trend is close to linear. Combining this result with the linear dependence of diffusiophoresis on the gradient strength, controlled through $c_p(0)$, shows that under our experimental conditions a combined mechanism of depletion-induced chemiphoresis and electrophoresis agrees with the observed profile in \partFig{fig:2}{c}.

We compare the experimentally measured $\UDP$ to Smoluchowski's theory (\Eq{eq:Smoluchowski}), with the calculated $\Delta \varphi$ shown in the inset of \partFig{fig:2}{c}. We again use $E \simeq -\Delta \varphi/L$ and $\varepsilon_r \simeq 80$, 
$\varepsilon_0\simeq\SI{8.85e-12}{\farad\per\metre}$,
$\zeta \simeq -\SI{68.5}{\milli\volt}$ (measured in Milli-Q water) and 
$\eta\simeq\SI{1}{\milli\pascal\second}$
which is reasonable for low molecular weight NaPSS at intermediate salt concentrations \cite{Gulati2023}. As shown in \partFig{fig:2}{c}, Smoluchowski's theory predicts much stronger electrophoresis compared to our experiments. 

To explain this mismatch we consider the charge density of the polyelectrolyte $c_p$. While NaPSS is a strong polyelectrolyte, meaning that the sulfonate groups are almost all fully ionized, the effective charge of a monomer $f$ is often below 1 due to Manning condensation, where counterions in the vicinity of the monomer groups reduce the effective electrostatic charge \cite{Manning1969}. The effective charge $f$ per monomer changes with $c_p$ and $c_s$, and its dependence is often nontrivial and nonlinear. A lower value of $f$ means the effective backbone charge density is smaller, which decreases the predicted electrophoretic velocity. We estimate $f$ by iteratively solving \Eqs{eq:BVP_our_system} with $c_{p}' = f \,c_p$ and adjusting $f$ until the velocity predicted from \Eq{eq:Smoluchowski} matches our experimental data. We find $f \simeq 0.1$--0.3, which is a reasonable value given our polyelectrolyte and salt concentration range \cite{Bordi2002,Bordi2004}. Therefore, it appears that counterion condensation can explain the mismatch between Smoluchowski's theory and our experiments.

\section{Particle Velocities in the Low-Salt Limit}

Interestingly, \partFig{fig:1}{g} suggests that under low salt conditions, which in our case corresponds to $c_s/c_p(0)<10^{-2}$, the steady state electrostatic potential across the channel saturates at $\Delta \varphi \simeq \SI{60}{\milli\volt}$, providing the maximum possible driving force for electrophoretic propulsion in the system. Furthermore, \partFig{fig:1}{g} shows that $c_p(0)$ has negligible influence on $\Delta \varphi$ in the limit where $c_s \rightarrow 0$, suggesting that electrophoresis becomes near-independent of polyelectrolyte concentration under low salt conditions.

To verify these predictions we repeated our experiments in Milli-Q water without adding salt. In \partFig{fig:3}{b} we observe an upward shift of $\UDP$ for all $c_p(0)$ at an added salt concentration $c_s = \SI{0}{\milli\molar}$ compared to the experiments at $c_s = \SI{150}{\milli\molar}$. This observation is consistent with predictions that charge effects are stronger at low salt. Furthermore, the slope in the \SI{0}{\milli\molar} salt experiments is slightly increased by a factor $\simeq 1.5$ compared to the \SI{150}{\milli\molar} salt systems. We note that the extrapolation of the fit for $c_p(0) \rightarrow 0$ to a significant finite value is nonphysical, as we do not expect any propulsion when $c_p(0)=0$. If $\UEP$ is approximately constant under low-salt conditions, as \partFig{fig:1}{g} suggests, it is possible that the slope under low salt conditions can be entirely attributed to a change in the chemiphoretic drift prefactor \cite{Sear2017}. Since this mechanism scales with polymer size, variations in the polyelectrolyte conformation, which change with concentration under low salt conditions \cite{Dobrynin2005}, could account for the observed dynamics.

\section{Propulsion in gradients of high MW polyelectrolyte}

As a final test, we examine the propulsion of colloids in higher molecular weight polyelectrolytes. Comparing NaPSS 70k to NaPSS 1M in \partFig{fig:3}{a} we find again a parabolic profile where the particles move down the polyelectrolyte gradient, while the velocities in the NaPSS 1M system are increased compared to the velocities in the NaPSS 70k system. We also show the diffusiophoretic velocities $\UDP$ in the NaPSS 1M system for multiple reservoirs concentrations $c_p(0)$, with and without added salt, in \partFig{fig:3}{c}. Interestingly, the propulsion velocities do not vary with $c_p(0)$, indicating that the dynamics has changed for these large polyelectrolytes. Furthermore, \partFig{fig:3}{c} shows that the velocities in the low salt system are increased compared to the \SI{150}{\milli\molar} system, which suggests that the dynamics are influenced by the Debye length $\lambda_D$, implying that the observed dynamics are related to electrostatics.

The lack of concentration dependence in \partFig{fig:3}{c} is predicted for diffusiophoresis of asymmetric electrolytes \cite{Gupta2019}, where local electric fields arising from the mismatch in diffusion coefficient and valency lead to a combined effect of locally induced electrophoresis and chemiphoresis, propelling the particles through the surrounding fluid \cite{Gupta2019}. For such systems, $\UCP \sim {\partial \ln c}/{\partial c} = (1/c)\,\partial c/\partial x$, with $c$ the ion concentration in the bulk, that satisfies electroneutrality. Since $(1/c)\,\partial c/\partial x$ has the order of magnitude $ \sim 1/L$, this means that the diffusiophoretic velocity for such systems does not depend on the absolute concentration of the (poly)electrolyte, as observed. 

The results presented in this section raise the question how local charge-induced diffusiophoresis could occur for NaPSS 1M but not for NaPSS 70k. In the steady state of the experiment, the charge distribution over the channel is expected to be similar for both polyelectrolytes. However, for NaPSS 1M we expect that the local diffusion coefficient is much slower compared to NaPSS 70k, due to the size and shape of the polymer, and possible entanglement effects. When the polyelectrolyte moves much slower, this can lead to stronger local dipoles that increases the electrical force on the particle surface, which can then become the dominant propulsion mechanism of the particles. We suggest that a detailed investigation of these mechanisms are an important avenue for future work.

\section{Summary and Conclusions}



In this work, we demonstrate that a macroscopic electric field emerges when charged species of different mobility are driven out of equilibrium. Using polyelectrolyte gradients as our model system, we show experimentally that the electric field leads to increased propulsion velocity of charged colloidal particles compared to neutral polymer gradients. While it is well established that macroscopic electric fields can form under equilibrium conditions \cite{Racsa2004}, our results suggest that electric fields can also emerge in non-equilibrium contexts, such as during evaporation of polymer solutions \cite{Huisman2025}.

Starting from a theoretical description of the liquid-junction problem, we predicted a linear polyelectrolyte concentration gradient along the channel, which we confirmed in experiments. The model also predicts that a macroscopic electric field is established across the channel, as there is a constant electrical current by the fluxes of the charged species, without a return path. The electric field then arises to balance the ion current, and increases the propulsion speeds of colloidal particles compared to a neutral polymer gradient because electrophoretic propulsion and depletion-induced chemiphoresis are additive contributions to the dynamics. 

Our observations gain additional significance because our experiments were performed at conditions representative of systems in industry and biological fluids, where polymer concentrations in the range 1--10\,wt\%
and salt concentrations in the range 0--\SI{150}{\milli\molar}
are often found. Therefore, we propose that a macroscopic electric field induced by polyelectrolyte gradients may be a universal feature in systems where polyelectrolytes are driven out of equilibrium. 

Finally, we showed that the propulsion velocity of microparticles in gradients of high molecular weight polyelectrolytes does not scale with gradient size, which was predicted for local charge-induced diffusiophoresis \cite{Gupta2019}. This opens the possibility that there is a transition between different phoretic regimes, that depends on the polymer molecular weight, which could be further investigated in future studies. 

Our findings, combining predictive theory with carefully controlled experiments, provide fundamental insight and serve as a platform for future research into particle dynamics in macromolecular gradients, for instance related to virus positioning in evaporating respiratory droplets or to the distribution of charged colloids in stratifying paint layers.

\section{Methods}

\subsection{Numerical calculations}

Our system of equations in \Eqs{eq:BVP_our_system} was numerically solved in python. Example codes are provided in the Supporting Information (SI). The SI also contains Refs.~\cite{Newman2021, Warren2020, Williams2020}.

\subsection{Experimental details}
\subsubsection*{Materials} 
Poly(ethylene glycol) (PEG, $M_{\rm W} =35$ kDa), poly(sodium 4-styrenesulfonate) (NaPSS, $M_{\rm W} \approx 70$ kDa), poly(sodium 4-styrenesulfonate) ($M_{\rm W} \approx 1,000$ kDa), sodium chloride (NaCl) and \ce{D2O} were purchased from Sigma-Aldrich. Polystyrene colloids (\ce{-COOH} endgroups, $R = \SI{1.25}{\micro\metre}$) were synthesized in-house.
\subsubsection*{Experimental setup}
Our experiments were performed using Ibidi $\upmu$-Slide Chemotaxis cells. The dimensions of the channel are 
\SI{1}{\milli\metre}\,$\times$\,\SI{2}{\milli\metre}\,$\times$\,\SI{70}{\micro\metre}
(length\,$\times$\,width\,$\times$\,height). We initialized the system such that the salt concentration and tracer particle density 
($2\times10^{-5}$\,\%\,w/w)
were homogeneous throughout the system. At the start of the experiment one of the two reservoirs contains the protein, while the other reservoir and the connecting channel did not. After setting up the experiment, we waited for 60 minutes before taking measurements to ensure that the system was in the steady state. 
\subsubsection*{Density matching}
Polymer and salt reservoirs were density matched to minimize density-driven flow and the associated particle motion in the channel. Because the polymer solution is intrinsically denser than the salt solution, we first measured the density of the polymer reservoir gravimetrically. Aliquots of fixed volume (\SI{1000}{\micro\litre}) were dispensed using a positive-displacement pipette, and each sample was weighed on an analytical balance (10 independent measurements per solution). The polymer density was calculated as $\rho = m/V$. The density of the salt reservoir was then adjusted by adding heavy water (70\% \ce{D2O}) until it matched the measured polymer density.

\subsection{Confocal Intensity Measurements and Particle Tracking}

\subsubsection*{Rhodamine B staining of PSS}
To enable fluorescence-based measurements of the polymer concentration field, poly(sodium 4-styrenesulfonate) (PSS, $M_W = \SI{70}{\kilo\dalton}$) was labeled non-covalently with rhodamine B which associates with PSS via $\pi$--$\pi$ interactions under acidic conditions; this was achieved by adjusting the PSS solution to pH\,3 prior to staining. Rhodamine B was added to the PSS solution at a molar ratio of approximately 1\,:\,10\,000 (dye\,:\,polymer) and mixed thoroughly before loading into the microfluidic device.

\subsubsection*{Confocal intensity measurements (2D imaging)}
Fluorescence intensity measurements of the polymer gradient were acquired on a Nikon A1R confocal microscope using a $4\times$ objective. For each condition, a single 2D confocal image ($xy$) of the channel was collected in Galvano scanning mode at $2048 \times 2048$ pixels. Polymer concentration profiles were obtained by extracting rhodamine fluorescence intensity along the channel axis. Intensity was averaged across the channel width within a rectangular region of interest to generate a one-dimensional intensity profile, $I(x)$, which was used as a proxy for the local polymer concentration.

\subsubsection*{3D confocal imaging for particle tracking}
Particle transport measurements were performed on the Nikon A1R confocal microscope using a $60\times$ water-immersion objective. Image sequences were acquired at $512 \times 512$ pixels at a frame rate of \SI{30}{\per\second}, corresponding to a lateral sampling of \SI{409}{\nano\metre} per pixel. For 3D tracking, time-resolved volumetric datasets were acquired as $z$-stacks with \SI{1}{\micro\metre} step size and at least 70 optical sections per stack (total stack depth $\geq\SI{70}{\micro\metre}$), spanning the measurement region in the channel.

\subsubsection*{3D particle tracking and trajectory analysis}
Three-dimensional particle tracking was performed using \texttt{trackpy} v0.5 following the standard 3D workflow. Particles were detected in each 3D frame using \texttt{trackpy}'s 3D feature-finding routines, and detected features were linked across time into trajectories using nearest-neighbor linking with a maximum displacement constraint. Particle positions were converted from pixels and $z$-step indices to physical units using the microscope calibration, and velocities were computed from finite differences of the tracked 3D positions.

\section*{Author Contributions}

M.H., A.A. and D.J.K. designed research; A.A., M.H. and D.J.K. performed experiments and analyzed data; P.B.W. and M.H. performed modeling and interpreted data; M.H. and D.J.K. wrote the first draft, and all authors reviewed and edited the manuscript.

\section*{Acknowledgments}

We thank Wilson C.K. Poon and Simon Titmuss for insightful discussions during the initial stages of the project. M.H., A.A. and D.J.K. acknowledge funding from the Dutch Research Council through grant OCENW.XS23.1.164.

\bibliography{selected}

\end{document}


\title{SUPPORTING INFORMATION\\[6pt]%
Self-Generated Electric Fields in Polyelectrolyte Gradients Increase Microparticle Transport}%

\author{Max Huisman}
    \thanks{These authors contributed equally to this work.}
    \affiliation{\lu}
\author{Ali Azadbakht}
    \thanks{These authors contributed equally to this work.}
    \affiliation{\lu}
\author{Patrick B. Warren}
    \affiliation{\ue}
    \affiliation{\hc}
\author{Daniela J. Kraft}
    \affiliation{\lu}

\date{\today}

\maketitle


\onecolumngrid

\newpage

\section{Derivation of the Boundary Value Problem}

Our experiment is a `restricted-diffusion junction' problem \cite{Newman2021}. To obtain a steady state solution, we start from the Nernst-Plack equations \cite{Newman2021,Warren2020}
\begin{equation}
    \label{eq:Nernst_Planck}
    \frac{-J_i}{D_i} = \frac{\partial c_i}{\partial x} + z_i c_i \frac{\partial \varphi}{\partial x}\,,
\end{equation}
where $J_i$ is the molar flux of species $i$, $D_i$ the diffusion coefficient [\unit{\metre\squared\per\second}], $z_i$ the charge valency, $c_i$ the charge concentration in [\unit{\mol\per\metre\cubed}] and $\varphi$ the electrostatic potential in [$k_B T/e$]. These equations are complemented by the charge neutrality condition
\begin{equation}
    \label{eq:charge_neutrality}
    \Sigma_i z_i c_i = 0\,.
\end{equation}
For the steady-state situation the fluxes $J_i$ are all constants that are determined by the boundary conditions. In this case, we can multiply \Eq{eq:Nernst_Planck} by $z_i$ and use the charge neutrality condition \Eq{eq:charge_neutrality} to obtain
\begin{equation}
    \label{eq:step1}
    -\Sigma_i z_i \frac{J_i}{D_i} = \left(\Sigma_i z_i^2c_i \right)\frac{\partial \varphi}{\partial x}.
\end{equation}
Defining the constant $J_i' = {J_i}/{D_i}$, we obtain from \Eqs{eq:Nernst_Planck} and~\ref{eq:step1} the system of ODEs
\begin{subequations}
    \label{eq:BVP}
\begin{equation}
    \frac{\partial \varphi}{\partial x} = \frac{-\Sigma_i z_i J_i'}{\Sigma_i z_i^2c_i}\,, ~~~ \frac{\partial c_i}{\partial x} = -J_i' - z_i c_i \frac{\partial \varphi}{\partial x}\,.
    \tag{\ref*{eq:BVP}a, \ref*{eq:BVP}b}
\end{equation}
\end{subequations}
\Eqs{eq:BVP} represents a boundary value problem (BVP) which can be readily solved using the SciPy \verb:solve_bvp: library in python. An example code is provided below.

For our system we consider the distribution of the polyelectrolyte NaPSS and the salt NaCl, leading to charge valencies for the polyelectrolyte monomer and co-ions $z_p = z_s = -1$ and $z_c = 1$ for the counterions. We define the charge distribution of the polyelectrolyte monomers, $c_p(x)$, and the co-ions, $c_s(x)$, both in \unit{\mole\per\metre\cubed}. The charge neutrality condition in \Eq{eq:charge_neutrality} obviates the need to consider the counterion distribution as an independent variable, as it is given by $c_p(x)+c_s(x)$. Similarly, we use the condition of no net ion current, $I = \Sigma_i z_i J_i = 0$, to eliminate the flux $J_c$ in favour of $J_p$ and $J_s$. Including the electrostatic potential $\varphi(x)$, from \Eqs{eq:BVP} our BVP consists of three ODEs,
\begin{subequations}\label{eq:BVP_our_system}
\begin{align}
    \frac{\partial \varphi}{\partial x} & = \frac{1}{2(c_p + c_s)} \left[\left(1-\frac{D_p}{D_c} \right)J'_p +\left(1-\frac{D_s}{D_c} \right)J'_s \right],  \\
    \frac{\partial c_p}{\partial x} &= -J_p' + c_p \frac{\partial \varphi}{\partial x}\,,~~~ 
    \frac{\partial c_s}{\partial x} = -J_s' + c_s \frac{\partial \varphi}{\partial x}\,, \tag{\ref*{eq:BVP_our_system}b, \ref*{eq:BVP_our_system}c}
\end{align}
\end{subequations}
with two unknown parameters $J_p'$ and $J_s'$. 

\section{Concentration Profiles in the Channel}

\begin{figure*}[h!]
\includegraphics[width=0.85\textwidth]{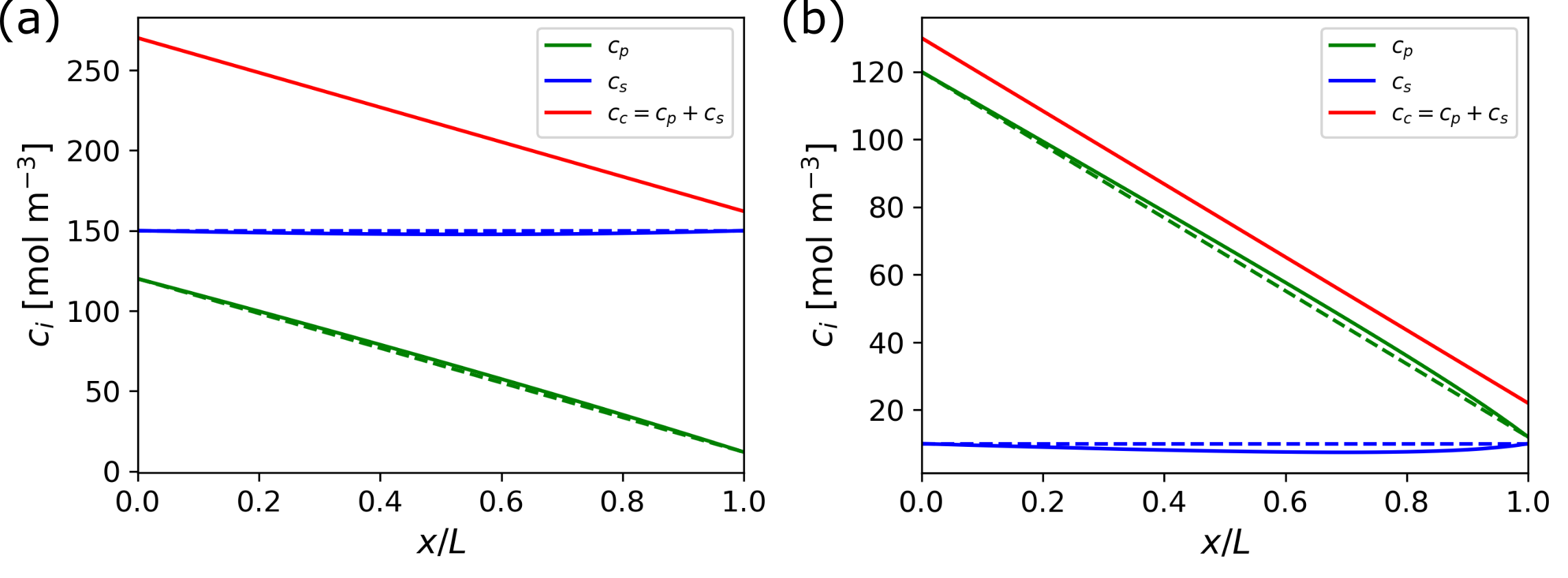}
\caption{\label{fig:s2} Steady-state concentration profiles calculated using the BVP presented in the main manuscript, where the polyelectrolyte reservoir contains $c_p(0)=\SI{120}{\mole\per\metre\cubed}$ and the background salt concentration is $c_s=\SI{150}{\milli\molar}$ (subfigure (a)) and $c_s=\SI{10}{\milli\molar}$ (subfigure (b)). Striped lines are visual aids for a linear profile.} 
\end{figure*}

Using the code snippet in \Sec{sec:BVP_code} we solved the boundary value problem presented above. The resulting concentration profiles in the channel are presented in \Fig{fig:s2}.

\section{Exact Solution to the BVP}

An exact solution to the BVP can be obtained by realizing that $c_p + c_s$ takes a linear profile. We can therefore write the \ansatz\
\begin{equation}
    c_p + c_s = a + b \xi\,, \label{eq:linear_Ansatz}
\end{equation}
where $\xi = x/L$ and $L$ is the length of the channel.

Next, we define for the first of Eqs. 5 in the main manuscript that
\begin{equation}
    \frac{\partial \varphi}{\partial x} = \frac{1}{(c_p + c_s)} K,
\end{equation}
with
\begin{equation}
    K = \frac{1}{2} \left[\left(1-\frac{D_p}{D_c} \right)J'_p +\left(1-\frac{D_s}{D_c} \right)J'_s \right]\,. \label{eq:K}
\end{equation}
Summing the second and third of Eqs. 5 and realizing that $\partial (c_p + c_s)/\partial \xi = b$ we obtain
\begin{equation}
    b = -J'_p -J'_s + K\,. \label{eq:help}
\end{equation}

Absorbing factors of $L$ into $J'_p$ and $J'_s$, the first two equations in Eqs. 5 in the main manuscript can be rewritten as
\begin{subequations}\label{eq:BVP_exact}
\begin{equation}
    \frac{\partial \varphi}{\partial \xi} = \frac{K}{a+b\xi}\,,~~~ 
    \frac{\partial c_p}{\partial \xi} + J'_p = c_p \frac{K}{a+b\xi}\,. \tag{\ref*{eq:BVP_exact}a, \ref*{eq:BVP_exact}b}
\end{equation}
\end{subequations}
Integration of the first of \Eqs{eq:BVP_exact} and using the boundary condition $\varphi(0)=0$ leads to
\begin{equation}
    \varphi = \frac{K}{b} \ln \left(1+\frac{b}{a} \xi \right).
\end{equation}
The second can then be integrated, for instance using an integrating factor, to obtain:
\begin{equation}
     c_p = \frac{J_p'}{J_p' + J_s'} (a+b\xi) + A (a+b\xi)^{K/b}, \\
\end{equation}
where $A$ is an integration constant. Using \Eq{eq:linear_Ansatz} we are given also
\begin{equation}
     c_s = \frac{J_s'}{J_p' + J_s'} (a+b\xi) - A (a+b\xi)^{K/b}. \\
\end{equation}

Now, we know from \Eq{eq:linear_Ansatz} that $a = c_s(0) + c_p(0)$ and thus also that $b = [c_s(1) + c_p(1)] - [c_s(0) + c_p(0)]$. Then, the unknown parameters $A$, $J_p'$ and $J_s'$ can be obtained by solving a non-linear equation set corresponding to the boundary conditions on $c_p$ (for example), supplemented by an additional constraint obtained by eliminating $K$ from \Eqs{eq:K} and~\ref{eq:help}. Solving this problem, for which we provide the code snippet in \Sec{sec:exact_code}, leads to the same concentration profiles presented in \Fig{fig:s2}.

\newpage
\section{Buoyancy driven flows}

\begin{figure*}[h!]
\includegraphics[width=0.6\textwidth]{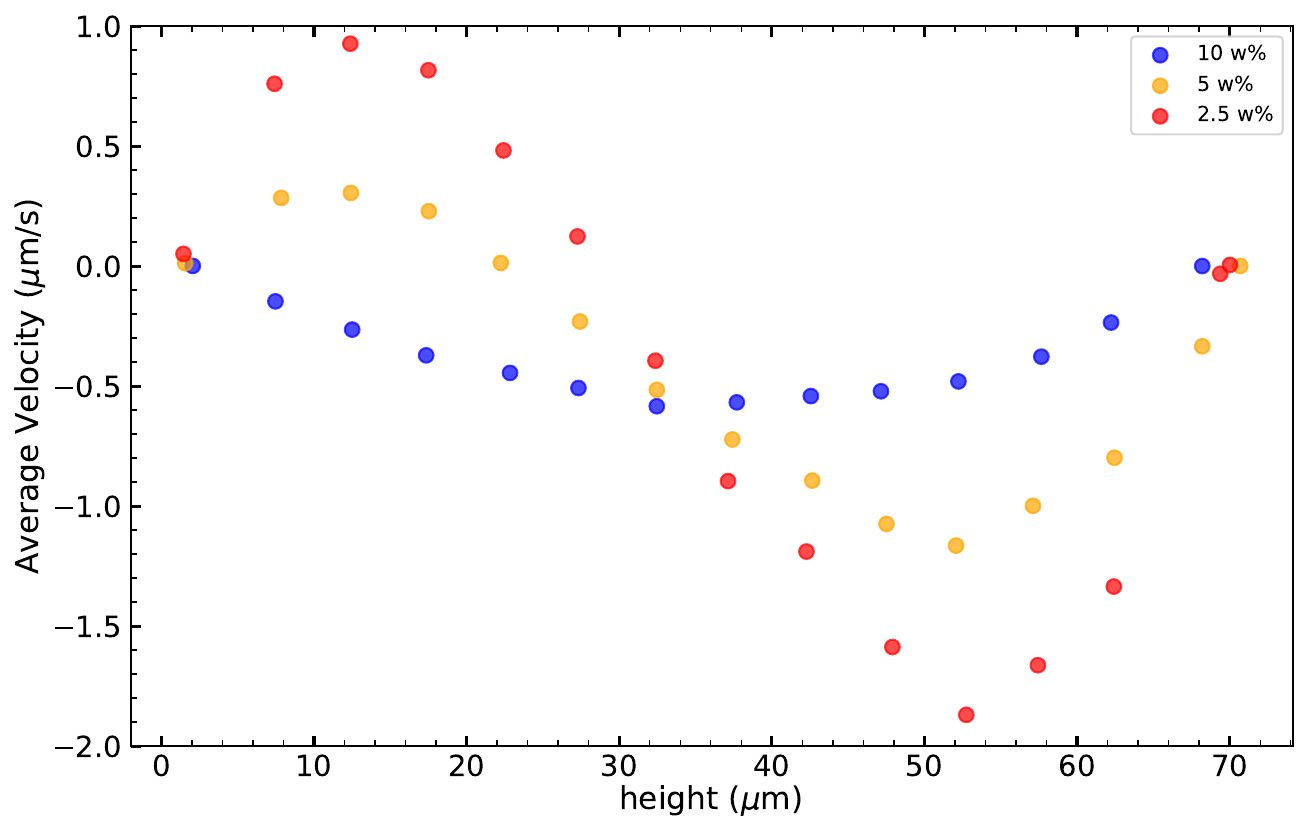}
\caption{\label{fig:s1} Velocity profile as a function of height for PSS~70k ($M_W=\SI{70}{\kilo\dalton}$), where the left reservoir has a density of \SI{1.05}{\gram\per\centi\metre\cubed} in \SI{0.15}{\molar} NaCl.} 
\end{figure*}

\Fig{fig:s1} shows the velocity profiles as a function of height where we did not density match the system. Clearly, at lower polymer concentrations we observe a backflow towards the high-concentration reservoir. Such a flow profile is similar to that observed in \cite{Williams2020}, which is attributed to a buoyancy-driven convective flow. 

\newpage
\section{Comparing Particle Velocities to Polymer Flux}

In the steady state the flux of the polyelectrolyte follows standard Fickian diffusion with $J = -D\, {\partial c}/{\partial x}$, meaning that the average velocity in the channel is
\begin{equation}
    \Upol = -\,\frac{D}{\rho}\,\frac{\partial c}{\partial x}\,,
\end{equation}
with $D$ the diffusion coefficient of the polyelectrolyte and $\rho$ its density. Assuming the linear concentration profile we estimate that in the steady state
\begin{equation}
    \Upol \approx -\frac{D}{\rho}\frac{\Delta c}{L}\,.
\end{equation}
Using $D = \SI{7e-11}{\metre\squared\per\second}$, 
$\rho = \SI{1400}{\kilo\gram\per\metre\cubed}$, 
$\Delta c = \SI{25}{\kilo\gram\per\metre\cubed}$,
and channel length $L = \SI{1e-3}{\metre}$
we obtain $\Upol\simeq \SI{1.25}{\nano\metre\per\second}$.
Clearly, this mass flux is much slower than the movement of the colloids with $U\sim\SI{1}{\micro\metre\per\second}$, which are therefore not pulled through the channel by the diffusing polymer.

\bibliography{selected}

\newpage
\section{Code snippet 1: solutions to the BVP}
\label{sec:BVP_code}

\lstset{
  language=Python,
  basicstyle=\ttfamily\scriptsize,
  lineskip=-1pt,
  keywordstyle=\color{blue},
  stringstyle=\color{red},
  commentstyle=\color{gray},
  numbers=left,
  numberstyle=\tiny,
  breaklines=true
}

\begin{lstlisting}
#!/usr/bin/env python3
import numpy as np
import matplotlib.pyplot as plt
from scipy.integrate import solve_bvp
from numpy import log as ln

Dp, Ds, Dc = 0.0, 2.03, 1.33 # diffusion coefficients: polymer, Cl-, Na+

cpI, cpII = 1.0, 0.0 # polymer gradient limits (reservoir values)
csI, csII = 0.1, 0.1 # salt gradient <-- adjust for plots in fig S1

def odes(xi, y, p): # ODEs as in eqs S5
    phi, cp, cs, Jp, Js = y[0], y[1], y[2], p[0], p[1]
    gradphi = ((1-Dp/Dc)*Jp + (1-Ds/Dc)*Js) / (2*cp + 2*cs)
    return np.vstack((gradphi, -Jp+cp*gradphi, -Js+cs*gradphi))

def bcs(ya, yb, p): # boundary conditions
    phi0, cp0, cs0 = ya[0], ya[1], ya[2]
    phi1, cp1, cs1 = yb[0], yb[1], yb[2]
    return np.array([phi0, cp0-cpI, cp1-cpII, cs0-csI, cs1-csII])

xi0 = np.linspace(0, 1, 5) # initial coarse grid, for starting guess
phi0 = np.zeros_like(xi0) # zero electric field, d(phi)/dx = 0
cp0 = cpI + (cpII-cpI)*xi0 # linear gradient , d(cp)/dx = cpII - cpI
cs0 = csI + (csII-csI)*xi0 # also a linear gradient
Jp0, Js0 = cpI-cpII, csI-csII # to correspond to the above solution

y0, p0 = np.vstack((phi0, cp0, cs0)), np.array([Jp0, Js0]) # initialise
res = solve_bvp(odes, bcs, xi0, y0, p=p0) # solve BVP
xi = np.linspace(0, 1, 41) # xi array for plotting purposes
phi, cp, cs = res.sol(xi) # extract solution interpolated onto xi

plt.plot(xi, phi, 'k-', label='phi')
plt.plot(xi, cp, 'g-', label='cp')
plt.plot(xi, cs, 'b-', label='cs')
plt.plot(xi, cp+cs, 'r-', label='cp+cs')
plt.plot(xi0, cp0, 'g--', label='cp(linear)') # re-use starting guess
plt.plot(xi0, cs0, 'b--', label='cs(linear)') # -- ditto --
plt.xlim(0, 1) ; plt.ylim(0, 1.8) # <-- adjust for different plots
plt.legend() ; plt.xlabel("x/L") ; plt.show() # produces plot as in fig S1
\end{lstlisting}

\newpage
\section{Code snippet 2: exact solution}
\label{sec:exact_code}

\begin{lstlisting}
import numpy as np
import matplotlib.pyplot as plt
from scipy.optimize import root
from numpy import log as ln

Dp, Ds, Dc = 0.0, 2.03, 1.33 # diffusion coefficients

cpI, cpII = 1.0, 0.0 # reservoir values, as before
csI, csII = 0.1, 0.1 # adjust as appropriate

a = cpI + csI # from eq S6
b = (cpII + csII) - (cpI + csI) # and likewise

def func(x): # define non-linear equation set for root finding
    A, Jp, Js = x # extract variables
    K = ((1-Dp/Dc)*Jp + (1-Ds/Dc)*Js)/2 # eq S8
    cp0 = Jp/(Jp+Js)*a + A * a**(K/b) # from eq S12 at xi = 0
    cp1 = Jp/(Jp+Js)*(a+b) + A * (a+b)**(K/b) # -- ditto --, at xi = 1
    return ([cp0-cpI, cp1-cpII, b+Jp+Js-K]) # boundary conditions

x0 = [0, cpI-cpII, csI-csII] # initial guess (linear profiles)
sol = root(func, x0) # solve the non-linear equation set
A, Jp, Js = sol.x # extract the solution
K = ((1-Dp/Dc)*Jp + (1-Ds/Dc)*Js)/2 # eq S8

xi = np.linspace(0, 1, 41) # array for plotting purposes
cp = Jp/(Jp+Js)*(a+b*xi) + A * (a+b*xi)**(K/b) # from eq S12
cs = Js/(Jp+Js)*(a+b*xi) - A * (a+b*xi)**(K/b) # from eq S13
phi = (K/b)*ln(1 + b*xi/a) # from eq S11

plt.plot(xi, phi, 'k-', label='phi')
plt.plot(xi, cp, 'g-', label='cp')
plt.plot(xi, cs, 'b-', label='cs')
plt.plot(xi, cp+cs, 'r-', label='cp+cs')
plt.plot([0, 1], [cpI, cpII], 'g--', label='cp(linear)') # linear profile
plt.plot([0, 1], [csI, csII], 'b--', label='cs(linear)') # -- ditto --
plt.xlim(0, 1) ; plt.ylim(0, 1.8) # adjust as appropriate
plt.legend() ; plt.xlabel("x/L") ; plt.show() # produces plots as before
\end{lstlisting}


